\documentclass[conference]{IEEEtran}

\usepackage{graphicx}
\usepackage{enumerate}
\usepackage{cite}
\usepackage{bm}
\usepackage[dvipsnames]{xcolor}
\usepackage{amsmath,amssymb,amsthm}
\usepackage{dutchcal}
\usepackage{stfloats}
\usepackage{suffix}
\usepackage{mathtools}
\usepackage[font=footnotesize,skip=2pt]{caption}
\begin{document}

\title{\vspace{0mm}RSMA for Dual-Polarized Massive MIMO Networks: A SIC-Free Approach \vspace{-4mm}}

\author{\IEEEauthorblockN{Arthur S. de Sena\IEEEauthorrefmark{1}\IEEEauthorrefmark{2},
		Pedro H. J. Nardelli\IEEEauthorrefmark{2}, Daniel B. da Costa\IEEEauthorrefmark{1}, Petar Popovski\IEEEauthorrefmark{3},\\
		Constantinos B. Papadias\IEEEauthorrefmark{4}, Mérouane Debbah\IEEEauthorrefmark{1}}
	\IEEEauthorblockA{\hspace{3mm}
	 \IEEEauthorrefmark{1} Technology Innovation Institute, United Arab Emirates\hspace{3mm} \IEEEauthorrefmark{2} Lappeenranta-Lahti University of Technology, Finland\\
	 \IEEEauthorrefmark{3} Aalborg University, Denmark \hspace{3mm}
	 \IEEEauthorrefmark{4} The American College of Greece, Greece}
	{Emails: arthurssena@ieee.org, pedro.nardelli@lut.fi,  daniel.costa@tii.ae, petarp@es.aau.dk,} 
	\\{  cpapadias@acg.edu, merouane.debbah@tii.ae \vspace{-7mm}}
}

\maketitle
\begin{abstract}
    Aiming at overcoming practical issues of successive interference cancellation (SIC), this paper proposes a dual-polarized rate-splitting multiple access (RSMA) technique for a downlink massive multiple-input multiple-output (MIMO) network. By modeling the effects of polarization interference, an in-depth theoretical analysis is carried out, in which we derive tight closed-form approximations for the outage probabilities and ergodic sum-rates. Simulation results validate the accuracy of the theoretical analysis and confirm the effectiveness of the proposed approach. For instance, under low to moderate cross-polar interference, our results show that the proposed dual-polarized MIMO-RSMA strategy outperforms the single-polarized MIMO-RSMA counterpart for all considered levels of residual SIC error.\vspace{-1mm}
\end{abstract}

\begin{IEEEkeywords}
	Massive MIMO, dual-polarized antenna arrays, rate-splitting multiple access.\vspace{-2mm}
\end{IEEEkeywords}


\section{Introduction}
Massive multiple-input multiple-output (MIMO) has become an indispensable technology for fifth-generation (5G) wireless communications systems and beyond.
Nevertheless, the installation of tens to hundreds of antennas in a tight physical space can create a strong channel correlation. This issue limits the size of practical antenna arrays and can hamper the performance of massive MIMO \cite{ref1}. Fortunately, the polarization domain provides an efficient way to mitigate this limitation. Specifically, since electromagnetic waves with orthogonal polarizations propagate with a low correlation, it is possible to implement orthogonal dual-polarized antennas and build an array with twice the number of antennas of a single-polarized array using identical physical dimensions \cite{ni3}. Moreover, the polarization domain offers a new degree of freedom (DoF) to MIMO systems, which can be exploited for generating multiplexing and diversity gains 
\cite{ni3,ref1, SenaT21}.

Efficient multiple access (MA) techniques are also essential for supporting the stringent requirements of beyond-5G systems. In particular, rate-splitting multiple access (RSMA) has arisen as a robust next-generation MA technique for MIMO systems \cite{Mao18, Dizdar21}. At the base station (BS), RSMA divides each users' message into two parts. The first part of each message is encoded into a single super symbol, called the common message, and mapped to the BS antennas through a common precoder, which is intended for all uses. The second part of each message, called private message, is transmitted via a private precoder that should be decoded only at the intended user. Upon reception, users rely on successive interference cancellation (SIC) to recover the transmitted messages.
Thanks to these features, RSMA can deliver high spectral and energy efficiencies, optimality in terms of DoF, and robustness even in scenarios with imperfect channel state information (CSI) \cite{Mao18}. Moreover, RSMA can outperform all conventional MA techniques, including non-orthogonal multiple access (NOMA) and orthogonal multiple access (OMA) techniques \cite{Dizdar21}.

Despite the above benefits, there are still issues that need to be studied and tackled. First, due to the SIC protocol in RSMA, the common message is always detected with interference from private messages, which has degrading effects on the system data rates. Moreover, existing RSMA-related works make the idealistic assumption that SIC can be carried out perfectly. However, due to hardware limitations, degraded CSI, and other issues, SIC errors are likely to happen in practice. As demonstrated in \cite{SenaISIC2020}, the residual interference left by imperfect SIC can strongly harm the performance of SIC-based schemes. Thus, strategies for combating the effects of imperfect SIC in RSMA are necessary. In particular, the performance superiority and additional DoF of dual-polarized MIMO systems can be exploited to alleviate interference issues of SIC \cite{SenaT21}. Nevertheless, to the best of our knowledge, the study of dual-polarized MIMO-RSMA systems is still missing in the literature, and there is no reported investigation of the harmful effects of imperfect SIC in RSMA schemes.

This major gap in the literature motivates the development of this work. Specifically, by modeling the practical issues of depolarization phenomena, we propose a low-complexity dual-polarized RSMA strategy for multiplexing common and private messages via the polarization domain in a massive MIMO network. This approach removes the need to execute SIC in the receivers and, consequently, frees the system from the detrimental effects of imperfect SIC. An in-depth theoretical study is carried out on the proposed transmission approach, where, first, we investigate the statistical properties of the achieved highly correlated channel gains. For overcoming the challenging statistical characterization, we assume that the channel gains are independent and determine their approximate distributions. Closed-form expressions for the outage probability and ergodic sum-rate are derived based on the obtained distributions. Simulation results supported by insightful discussions validate the theoretical analysis and demonstrate the potential performance improvements enabled by the dual-polarized RSMA scheme.



\vspace{.5mm}

\noindent  \textbf{\textit{Notation and Special Functions:}} Bold-faced lower-case letters denote vectors and upper-case represent matrices. The transpose and the Hermitian transpose of $\mathbf{A}$ are represented, respectively, by $\mathbf{A}^T$ and $\mathbf{A}^H$, the operator tr$\{\mathbf{A}\}$ computes the trace of $\mathbf{A}$, and $[\mathbf{A}]_{i:j}$ returns a sub-matrix of $\mathbf{A}$ containing its columns from $i$ to $j$. The symbol $\otimes$ represents the Kronecker product, $\mathbf{I}_M$ represents the identity matrix of dimension $M\times M$, and $\mathbf{0}_{M, N}$ denotes the $M\times N$ matrix with all zero entries. In addition, $\mathrm{E}(\cdot)$ denotes expectation, $\Gamma(\cdot)$ is the Gamma function \cite[eq. (8.310.1)]{ref8}, $\gamma(\cdot,\cdot)$ is the lower incomplete Gamma function \cite[eq. (8.350.1)]{ref8}, $\bm{e}_n(\cdot)$ denotes the truncated Taylor series of the exponential function with $n$ terms \cite[eq. (1.211.1)]{ref8}, and $\mathrm{Ei}(\cdot)$ corresponds to the exponential integral \cite[eq. (8.211.1)]{ref8}.

\section{System Model}

\begin{figure}[t]
	\centering
	\includegraphics[width=.78\linewidth]{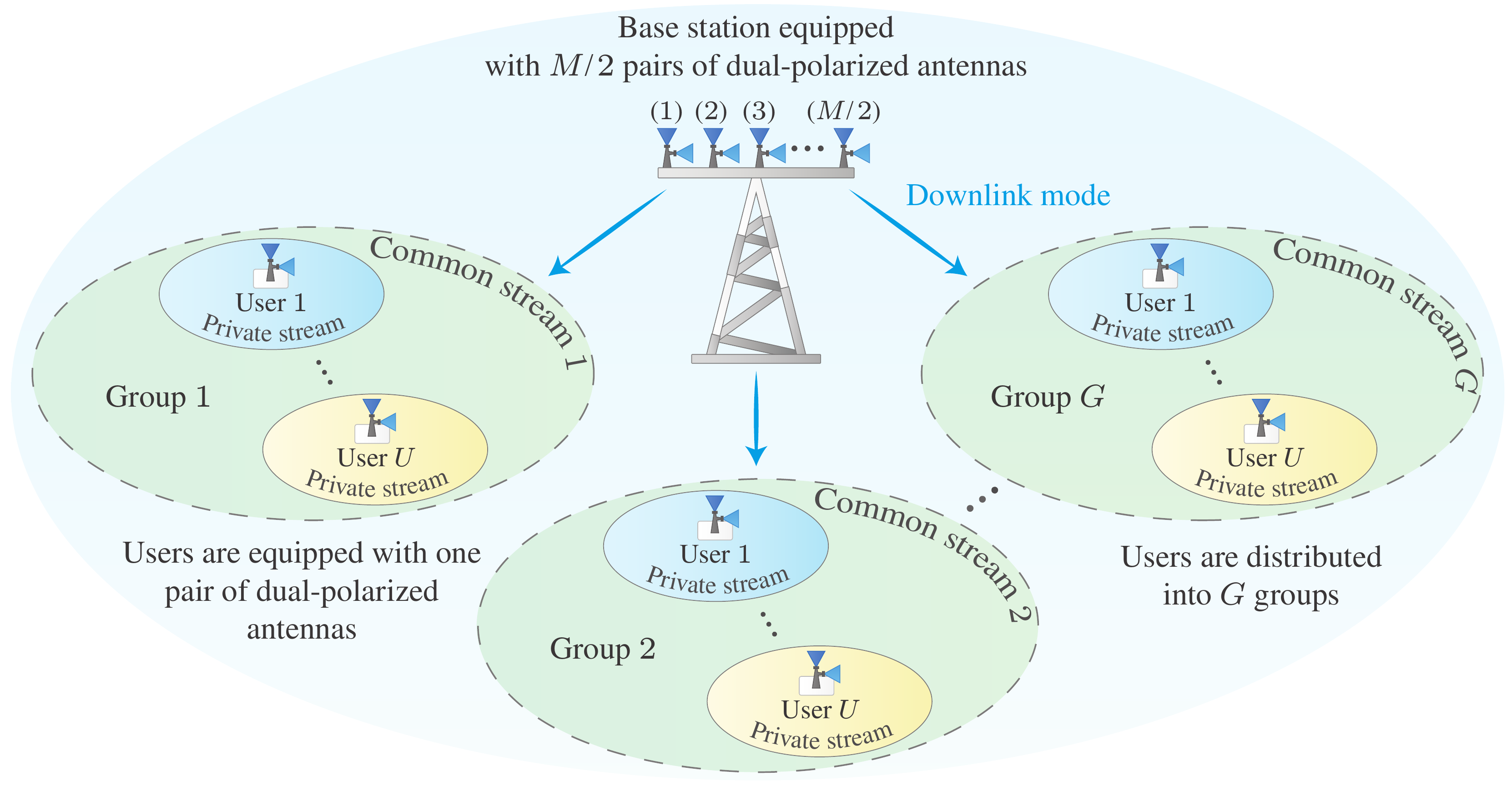}
	\caption{Proposed system model. A dual-polarized massive MIMO-RSMA base station serves dual-polarized users distributed into different spatial groups.}\label{f1}
\end{figure}

We consider a massive MIMO network in which a single BS communicates in downlink mode with $L$ users. The BS is equipped with $M/2$ pairs of co-located dual-polarized transmit antennas, and each user employs one pair of dual-polarized receive antennas, such that each antenna pair contains one horizontally and one vertically polarized antenna element.
Due to the closely spaced antennas, the wireless channels of users located in similar angular sectors become correlated.
The BS exploits this characteristic and clusters the users into $G$ groups based on the likeness of their channel covariance matrices. For simplicity, we assume that each group contains $U$ users, i.e., $L = G U$, and that users within each group share a common covariance matrix $\mathbf{\bar{R}}_g = \mathbf{I}_{2} \otimes \mathbf{R}_g$, in which $\mathbf{R}_g$ is the covariance matrix corresponding to each polarization with rank denoted by $r_g$. Under such assumptions, we can represent the wireless channel for the $u$th user in the $g$th group by \cite{ni3}
\begin{align}\label{eq01}
    \mathbf{H}_{gu} &= \sqrt{\zeta_{gu}}\begin{bmatrix} \mathbf{h}^{vv}_{gu} & \sqrt{\chi}\mathbf{h}^{vh}_{gu} \\
    \sqrt{\chi}\mathbf{h}^{hv}_{gu} & \mathbf{h}^{hh}_{gu}\end{bmatrix} \nonumber\\
     &=  \begin{bmatrix} \sqrt{\zeta_{gu}}\mathbf{U}_g\bm{\Lambda}^{\frac{1}{2}}_g\mathbf{g}^{vv}_{gu} & \sqrt{\zeta_{gu} \chi}\mathbf{U}_g\bm{\Lambda}^{\frac{1}{2}}_g\mathbf{g}^{vh}_{gu} \\ 
    \sqrt{\zeta_{gu} \chi}\mathbf{U}_g\bm{\Lambda}^{\frac{1}{2}}_g \mathbf{g}^{hv}_{gu} & \sqrt{\zeta_{gu}}\mathbf{U}_g\bm{\Lambda}^{\frac{1}{2}}_g\mathbf{g}^{hh}_{gu}\end{bmatrix} \hspace{-1mm} \in \mathbb{C}^{M\times 2}, \hspace{-1mm}
\end{align}
where $\bm{\Lambda}_g \in \mathbb{R}^{\bar{r}_g \times \bar{r}_g}_{>0}$ is a diagonal matrix formed by $\bar{r}_g$ nonzero eigenvalues of $\mathbf{R}_g$ sorted in descending order, $\mathbf{U}_g \in \mathbb{C}^{\frac{M}{2}\times \bar{r}_g}$ comprises the corresponding $\bar{r}_g$ left eigenvectors of $\mathbf{R}_g$ obtained from the singular value decomposition (SVD), $\mathbf{g}^{ij}_{gu} \in \mathbb{C}^{\bar{r}_g \times 1}$ is a vector that models the reduced-dimension fast-fading channels from polarization $i$ to polarization $j$, in which $i, j \in \{v, h\}$, with $v$ and $h$ denoting, respectively, the vertical and horizontal polarizations, whose entries follow the complex Gaussian distribution with zero-mean and unit-variance, $\zeta_{gu}$ denotes the large-scale fading coefficient, and $\chi$ represents the inverse of the cross-polar discrimination (iXPD) parameter
that models the level of cross-polar interference.

\subsection{RSMA for Dual-Polarized Massive MIMO}
Inspired by the recent works \cite{ni3, SenaISIC2020,ref1}, we implement a two-stage transmission approach where, in the first stage, spatial multiplexing is performed for separating the multiple groups of users and, in the second stage, the RSMA technique is employed to serve the users within each group. As anticipated in Section I, the BS splits each data message into a common and a private part. The common parts are encoded into a common symbol\footnote{RSMA schemes with multiple common symbols and multiple layers of SIC also exist \cite{Mao18}. However, the study of these more complicated strategies goes beyond the scope of this work.}, $c_g$, which should be decoded by all users within the $g$th group, while the private parts are encoded into private symbols, $p_{gu}$, each one intended to a particular user. Lastly, the private and common symbols are multiplied by linear precoders and then superimposed in the power domain for transmission, resulting in the following data stream:
\begin{align}\label{eqsig1}
    \mathbf{x} &= \sum_{m = 1}^{G}  \mathbf{K}_{m}\left( \mathbf{c}_{m} \sqrt{\alpha_{m}} c_{m} +  \sum_{n = 1}^{U} \mathbf{p}_{mn} \sqrt{\beta_{mn}} p_{mn} \right),
\end{align}
where $\alpha_{m}$ and $\beta_{mn}$ denote, respectively, the power allocation coefficients for the common and private messages, $\mathbf{K}_{m}  = \mathbf{I}_{2} \otimes \mathbf{F}_m \in \mathbb{C}^{M\times\bar{M}}$ is the precoding matrix responsable for performing spatial multiplexing of user groups, in which $\mathbf{F}_m \in \mathbb{C}^{\frac{M}{2}\times\frac{\bar{M}}{2}}$ represents the precoding matrix for each polarization, $\bar{M}$ is a parameter that controls the dimension of the projected channel, and $\mathbf{c}_{m} \in \mathbb{C}^{\bar{M} \times 1}$ and $\mathbf{p}_{mn} \in \mathbb{C}^{\bar{M} \times 1}$ are the precoding vectors for the common and private messages, respectively, satisfying $\|\mathbf{c}_{m}\|^2 = 1$ and $\|\mathbf{p}_{mn}\|^2 = 1$.

At the users' side, the common message is detected first while treating the private message as interference. Then, each user executes SIC to subtract the common message from the composite signal. After that, the private message is finally recovered. This conventional approach is effective for retrieving the transmitted messages and can deliver remarkable performance gains. However, as discussed before, there are limitations. 
In this paper, we exploit the polarization domain to cope with the SIC-related issues, as explained next.

Instead of transmitting a superimposed stream, the private and common messages are transmitted through independent polarized data streams, i.e., each message is assigned to one polarization. Such a strategy will enable users to decode common and private messages without relying on SIC. In addition to freeing the system from errors of imperfect SIC, users should be able to recover the common message without interference of the private messages, i.e., in ideal conditions. In practice, however, users will experience cross-polar interference, an issue that will be investigated in our analysis. For simplicity, the common message is assigned to the vertical polarization and the private messages to the horizontal polarization.
With this strategy, the precoding vector for the common message can be written as $\mathbf{c}_{g} = [(\mathbf{c}^v_{g})^H, \mathbf{0}_{1, \frac{\bar{M}}{2}}]^H $, and for the private message as $\mathbf{p}_{gu} = [ \mathbf{0}_{1, \frac{\bar{M}}{2}}, (\mathbf{p}^h_{gu})^H]^H $, where $\mathbf{c}^v_{m} \in \mathbb{C}^{\frac{\bar{M}}{2} \times 1}$ and $\mathbf{p}^h_{mn} \in \mathbb{C}^{\frac{\bar{M}}{2} \times 1}$ are the precoding vectors for the common and private messages corresponding to polarizations $v$ and $h$, respectively.
As a result, the $u$th user in the $g$th group will receive the following signal
\begin{align}\label{eq03}
    \mathbf{y}_{gu} &=  
    \mathbf{H}^H_{gu} \hspace{-1mm} \sum_{m = 1}^{G} \hspace{-.8mm} \mathbf{K}_{m} \hspace{-1mm}
    \begin{bmatrix}
    \mathbf{c}^v_{m} \sqrt{\alpha_{m}} c_{m} \\
    \sum_{n = 1}^{U} \mathbf{p}^h_{mn} \sqrt{\beta_{mn}} p_{mn}
    \end{bmatrix} +
    \begin{bmatrix}
    n^v_{gu} \\ n^h_{gu} 
    \end{bmatrix},
\end{align}
where $n^i_{gu} \in \mathbb{C}$ is the additive white noise observed in polarization $i \in \{v, h\}$, which follows the complex Gaussian distribution with zero mean and variance $\sigma^2$.

\section{Precoder Design}\label{secprec}
In order to employ RSMA to each group separately, the precoding matrix $\mathbf{K}_g$ should be designed to eliminate the inter-group interference.
Such a goal can be accomplished by exploiting the null space spanned by the left eigenvector matrices of interfering groups, i.e., by defining $ \mathbf{U}^{*}_{g} = [\mathbf{U}_{1},\cdots,\mathbf{U}_{g-1}, \mathbf{U}_{g+1}, \cdots, \mathbf{U}_{G}] \in \mathbb{C}^{\frac{M}{2} \times \sum_{g'\neq g} \bar{r}_{g'}}$, $\mathbf{F}_{g}$ can be constructed from the orthonormal basis of $\mathrm{null}\{\mathbf{U}^{*}_{g}\}$, which can be obtained from the eigenvectors corresponding to the zero eigenvalues of $\mathbf{U}^{*}_{g}$. More specifically, let $\mathbf{E}^{0}_{g} \in \mathbb{C}^{\frac{M}{2}\times \frac{M}{2} - \sum_{g'\neq g} \bar{r}_{g'}}$ denote the matrix comprising the last $\frac{M}{2} - \sum_{g'\neq g} \bar{r}_{g'}$ left eigenvectors of $\mathbf{U}^{*}_{g}$ computed via SVD. Then, the desired precoding matrix can be given by $\mathbf{K}_{g} = \mathbf{I}_{2} \otimes \mathbf{F}_{g} = \mathbf{I}_{2} \otimes \big[\mathbf{E}^{0}_{g}\big]_{1:\frac{\bar{M}}{2}}$, in which, due to the dimension of $\mathbf{E}^{0}_{g}$, it is required that $\frac{\bar{M}}{2} \leq \frac{M}{2} - \sum_{g'\neq g} \bar{r}_{g'}$, $\frac{M}{2} > \sum_{g'\neq g} \bar{r}_{g'}$, and $\frac{\bar{M}}{2} \leq \bar{r}_{g} \leq r_{g}$. To satisfy these constraints, the parameter $\bar{r}_g$ is configured as $\mathrm{min}\{r_g, \lfloor (\frac{M}{2} - \frac{\bar{M}}{2})/(G  -1)  \rfloor \}$.


Now, we concentrate on the design of the precoding vector for the private massages. The role of $\mathbf{p}^{h}_{gu}$ is to ensure that each private message reaches only its intended user in the assigned polarization. This implies that $\mathbf{p}^h_{gu}$ must be orthogonal to the subspace spanned by the effective channel
$(\mathbf{h}^{hh}_{gu'})^H\mathbf{F}_{g}$ of users $u' \neq u$, which can be achieved by
\begin{align}
    \mathbf{p}^h_{gu} = \mathrm{null}\{ \mathbf{F}^H_{g}[\mathbf{h}^{hh}_{g1}, \cdots, \mathbf{h}^{hh}_{g(u-1)}, \mathbf{h}^{hh}_{g(u+1)}, \cdots, \mathbf{h}^{hh}_{gU}] \},
\end{align}
in which we must have $\bar{M}> U - 1$ to ensure the existence of a non-trivial null space.

In turn, $\mathbf{c}^{v}_{g}$ should be designed to broadcast the common messages to all users. Different strategies and optimization procedures have been proposed in the literature, including matched filter precoding (MFP), weighted MFP, precoding for max-min fairness, among others \cite{Mao18}. However, such sophisticated precoding designs can lead to an intractable analysis. As an alternative, we construct $\mathbf{c}^{v}_{g}$ as a random precoder with independent and identically distributed entries following the standard complex Gaussian distribution, as in \cite{Dizdar21}.



\section{Performance Analysis}
In this section, we carry out a statistical characterization of the effective channel gains observed in the signal-to-interference-plus-noise ratios (SINRs), based on which we derive closed-form approximations for the outage probabilities. Tight approximations for ergodic sum-rates are also derived.

\subsection{Signal Detection and SINR Analysis}
With $\mathbf{K}_{g}$ and $\mathbf{p}^h_{gu}$ given in Section III, the signal in \eqref{eq03} can be simplified as
\begin{align}\label{detsgia1}
    &\mathbf{y}_{gu} =  
    \begin{bmatrix} \sqrt{\zeta_{gu}} (\mathbf{h}^{vv}_{gu})^H\mathbf{F}_{g}\mathbf{c}^v_{g} \sqrt{\alpha_{g}} c_{g} \\ \sqrt{\zeta_{gu}}
    (\mathbf{h}^{hh}_{gu})^H\mathbf{F}_{g} \mathbf{p}^h_{gu} \sqrt{\beta_{gu}} p_{gu} \end{bmatrix} \nonumber\\
    &+
    \begin{bmatrix} \sqrt{\zeta_{gu}\chi}(\mathbf{h}^{hv}_{gu})^H\mathbf{F}_{g} \sum_{n = 1}^{U} \mathbf{p}^h_{gn} \sqrt{\beta_{gn}} p_{gn} \\ \sqrt{\zeta_{gu}\chi}(\mathbf{h}^{vh}_{gu})^H\mathbf{F}_{g}\mathbf{c}^v_{g} \sqrt{\alpha_{g}} c_{g} \end{bmatrix}
    +
    \begin{bmatrix}
    n^v_{gu} \\ n^h_{gu} 
    \end{bmatrix}.
\end{align}

As can be observed in \eqref{detsgia1}, upon reception, users become able to recover the common and private messages directly from the assigned polarization, effortless without SIC. However, both messages are corrupted by cross-polar interference that $\mathbf{p}^h_{gu}$ is unable to cancel. Consequently, the SINR experienced when detecting the common message can be given by
\begin{align}\label{sinr_a1_c}
    \gamma^{c}_{gu} &= \frac{ \zeta_{gu} |(\mathbf{h}^{vv}_{gu})^H\mathbf{F}_{g}\mathbf{c}^{v}_{g}|^2 \alpha_{g}}
    { \zeta_{gu}\chi \sum_{n=1}^{U}|(\mathbf{h}^{hv}_{gu})^H\mathbf{F}_{g} \mathbf{p}^{h}_{gn}|^2 \beta_{gn} + \sigma^2}.
\end{align}

Analogously, the SINR observed by the $u$th user when decoding its private message can be written as
\begin{align}\label{sinr_a1_p}
    \gamma^{p}_{gu} = \frac{ \zeta_{gu} |(\mathbf{h}^{hh}_{gu})^H\mathbf{F}_{g} \mathbf{p}^h_{gu}|^2 \beta_{gu} }
    {\zeta_{gu}\chi|(\mathbf{h}^{vh}_{gu})^H\mathbf{F}_{g}\mathbf{c}^v_{g}|^2 \alpha_{g}  + \sigma^2}.
\end{align}\vspace{-3mm}

\subsection{Statistical Characterization of Channel Gains}\label{statc_a1}
Let the gain in the numerator of \eqref{sinr_a1_c} be denoted by $\varsigma^c_{gu} = \zeta_{gu} |(\mathbf{h}^{vv}_{gu})^H\mathbf{F}_{g}\mathbf{c}^{v}_{g}|^2 \alpha_{g}$ and the interference term by $\omega^c_{gu} = \zeta_{gu}\chi \sum_{n=1}^{U}|(\mathbf{h}^{hv}_{gu})^H\mathbf{F}_{g} \mathbf{p}^{h}_{gn}|^2 \beta_{gn}$. Then, the squared norm in $\varsigma^c_{gu}$ can be expanded as $|(\mathbf{h}^{vv}_{gu})^H\mathbf{F}_{g}\mathbf{c}^{v}_{g}|^2 =  (\mathbf{c}^{v}_{g})^H\mathbf{F}^H_{g}\mathbf{h}^{vv}_{gu}(\mathbf{h}^{vv}_{gu})^H\mathbf{F}_{g}\mathbf{c}^{v}_{g}
= \mathrm{tr}\{\mathbf{F}^H_{g}\mathbf{h}^{vv}_{gu}(\mathbf{h}^{vv}_{gu})^H\mathbf{F}_{g}\mathbf{c}^{v}_{g}(\mathbf{c}^{v}_{g})^H\}$.
Given that $\mathbf{c}^{v}_{g}$ is an isotropic unit vector independent of $(\mathbf{h}^{vv}_{gu})^H\mathbf{F}_{g}$ and that $\mathrm{E}\{\mathbf{c}^{v}_{g}(\mathbf{c}^{v}_{g})^H\} = \frac{1}{\bar{M}}\mathbf{I}_{\bar{M}}$, we have that $\mathrm{E}\{|(\mathbf{h}^{vv}_{gu})^H\mathbf{F}_{g}\mathbf{c}^{v}_{g}|^2 \} = \frac{1}{\bar{M}}\mathrm{tr}\{\mathbf{F}^H_{g}\mathbf{h}^{vv}_{gu}(\mathbf{h}^{vv}_{gu})^H\mathbf{F}_{g}\}$.
Since $\mathbf{g}^{vv}_{gu}$ in \eqref{eq01} is a complex Gaussian distributed vector and $\mathbf{F}_{g}$ is a semi-unitary matrix (with orthonormal columns), the vector $\mathbf{F}^H_{g}\mathbf{h}^{vv}_{gu}$ is still complex Gaussian distributed.
Consequently, the gain $\varsigma^c_{gu} = \zeta_{gu} |(\mathbf{h}^{vv}_{gu})^H\mathbf{F}_{g}\mathbf{c}^{v}_{g}|^2 \alpha_{g}$ follows a Gamma distribution with shape parameter $1$ and rate parameter
$\phi/\zeta_{gu}\alpha_{g}$, with $\phi = \bar{M}/\mathrm{tr}\{ \mathbf{F}^H_{g}\mathbf{R}_{g}\mathbf{F}_{g}\}$, 
whose probability distribution function (PDF) can be found in \cite[eq. (15.1.1)]{Krishnamoorthy2006}.

Now, we focus on $\omega^c_{gu}$. Note that $\omega^c_{gu}$ consists of a sum of $U$ correlated random variables (RVs). Consequently, determining the exact distribution of $\omega^c_{gu}$ becomes challenging. For overcoming this issue, we approximate the PDF of $\omega^c_{gu}$ by assuming that the sum terms are independent. Then, we can expand the squared norm in $\omega^c_{gu}$ as $\mathrm{tr}\{\mathbf{F}^H_{g}\mathbf{h}_{gu}^{hv}(\mathbf{h}_{gu}^{hv})^{H}\mathbf{F}_{g}\mathbf{p}^{h}_{gn}(\mathbf{p}^{h}_{gn})^H\}$. Since $\mathbf{p}^{h}_{gn}$ is uniformly distributed with the Haar measure $O(\bar{M})$, we have that $\mathrm{E}\{\mathbf{p}^{h}_{gn}(\mathbf{p}^{h}_{gn})^H\} = \frac{1}{\bar{M}}\mathbf{I}_{\bar{M}}$. As a result, the distribution of each term of the summation in $\omega^c_{gu}$ follows a Gamma distribution with shape parameter $1$ and rate parameter ${\phi}/{\zeta_{gu} \chi\beta_{gn}}$, for $n=1,\cdots,U$.
However, given that we employ a uniform power allocation among the private messages\footnote{As stated in \cite{Dizdar21}, uniform power allocation for multi-user MIMO is widely employed in the literature and used in practical 4G and 5G systems.}, we have that $\beta_{g1} = \beta_{g2} = \cdots = \beta_{gU}$. This implies that the terms of the sum in $\omega^c_{gu}$ have equal rate parameters. Consequently, the PDF of $\omega^c_{gu}$ can be approximated by a Gamma PDF with shape parameter $U$ and rate parameter ${\phi}/{\zeta_{gu} \chi\beta_{gu}}$. 

We characterize now the gains in the SINR for the private messages in \eqref{sinr_a1_p}. Let $\varsigma^{p}_{gu} = \zeta_{gu} |(\mathbf{h}^{hh}_{gu})^H\mathbf{F}_{g} \mathbf{p}^h_{gu}|^2 \beta_{gu}$ and $\omega^{p}_{gu} = \zeta_{gu}\chi|(\mathbf{h}^{vh}_{gu})^H\mathbf{F}_{g}\mathbf{c}^v_{g}|^2 \alpha_{g}$. As can be noticed, $\varsigma^{p}_{gu}$ and $\omega^{p}_{gu}$ have a similar form of $\varsigma^c_{gu}$, which implies that such gains can be characterized similarly as we have characterized $\varsigma^c_{gu}$. Therefore, the full details for this statistical characterization are omitted. In short, $\varsigma^{p}_{gu}$ and $\omega^{p}_{gu}$ follow Gamma distributions with shape parameters $1$ and rate parameters given by $\phi/\zeta_{gu}\beta_{gu}$ and $\phi/\zeta_{gu} \chi\alpha_{g}$, respectively. 
\vspace{-1.5mm}

\subsection{Outage Probability}

In RSMA, users need to successfully decode both common and private messages for them to be able to reconstruct the intended original information. Therefore, an outage event will occur either if the rate of the common message or the rate of the private message drops below the corresponding target data rate. Mathematically, the outage probability for the $u$th user in the $g$th group can be obtained by
\begin{align}\label{gen_out}
    P^{\text{out}}_{gu} &= P^c_{gu} \cup P^p_{gu} = P^c_{gu} + P^p_{gu} - P^c_{gu} P^p_{gu},
\end{align}
where $P^c_{gu} = \mathrm{Pr}\{\log_2(1 + \gamma^{c}_{gu}) < R^{c}_{g}\}$ and $P^p_{gu} = \mathrm{Pr}\{\log_2(1 + \gamma^{p}_{gu}) < R^{p}_{gu}\}$, with $R^{c}_{g}$ and $R^{p}_{gu}$ representing, respectively, the target data rates for the common and private messages, which are measured in bits per channel use (bpcu). A tight closed-form approximation for the outage probability of the proposed strategy is derived next.

\paragraph*{Proposition I} When the BS employs the proposed dual-polarized MIMO-RSMA scheme, the outage probability for the common message experienced by the $u$th user in the $g$th group can be approximated by
\begin{align}\label{out_comm_a1}
    P^c_{gu} &= 1 - \left(\frac{\alpha_{g}}{\alpha_{g} + \chi\beta_{gu}\tau_g^c } \right)^U  e^{-\frac{\phi\tau_g^c}{\rho\zeta_{gu}\alpha_{g}}},
\end{align}
where $\tau^{c}_{g} = 2^{R^{c}_{g}} - 1$ and $\rho = 1/\sigma^2$ is the signal-to-noise ratio (SNR).

\textit{Proof:} Please, see Appendix \ref{ap1}.  \hfill $\blacksquare$

\paragraph*{Proposition II} When the BS employs the proposed dual-polarized MIMO-RSMA scheme, the outage probability for the private message experienced by the $u$th user in the $g$th group can be approximated by
\begin{align}\label{out_private_a1}
    P^p_{gu} & = 1 - \frac{\beta_{gu}}{ \chi\alpha_{g}\tau_{gu}^p + \beta_{gu}} e^{-\frac{\phi \tau_{gu}^p}{\rho\zeta_{gu}\beta_{gu}}},
\end{align}
where $\tau_{gu}^p = 2^{R_{gu}^{p}} - 1$. 

\textit{Proof:} Please, see Appendix \ref{ap2}.  \hfill $\blacksquare$

The general outage probability, $P^{\text{out}}_{gu}$, for the dual-polarized MIMO-RSMA scheme can be easily obtained by replacing \eqref{out_comm_a1} and \eqref{out_private_a1} into \eqref{gen_out}. 
\vspace{-2mm}

\subsection{Ergodic Sum-Rates}
For ensuring a successful decoding at all users, the ergodic rate for the common message is computed by averaging the minimum of the common message's instantaneous rates achieved by the users \cite{Mao18}. Therefore, the ergodic sum-rate for the $g$th group can be computed by
\begin{align}\label{ergsm_g_rsma}
C_g &=  C^p_{g} + C^c_{g}, 
\end{align}
where $C^p_{g} = \mathrm{E}\{  \sum_{u=1} \hspace{-6mm}{}^{U} \hspace{4mm} \log_2(1 + \gamma^p_{gu})\}$ and $C^c_{g} = \mathrm{E}\{ \sum_{u=1} \hspace{-6mm}{}^{U} \hspace{4mm} \min_{l\in \{1, \cdots, U\} }\{ \log_2(1 + \gamma^c_{gl}) \} \}$ are the ergodic sum-rates for the private and common messages, respectively. The following propositions provide a tight approximation for the ergodic sum-rate.

\paragraph*{Proposition III} When the BS employs the proposed dual-polarized MIMO-RSMA scheme, the ergodic sum-rate for the common message experienced in the $g$th group can be approximated by
\begin{align}\label{erg_comm_a1}
    &C^c_g =  \text{\small $\sum_{u=1}^{U} \frac{(-1)^{U^2 - 1}}{\mathrm{ln}(2)}  \left(\frac{\alpha_{g}}{\chi\beta_{gu} - \alpha_{g}} \right)^{U^2} e^{\frac{\phi \sum_{l=1}^{U} \frac{1}{\zeta_{gl}} }{\rho \alpha_{g}} }$} \nonumber\\
    & \text{\small $\times \left[ \mathrm{Ei}\left(- \frac{\phi \sum_{l=1}^{U} \frac{1}{\zeta_{gl}} }{\rho \alpha_{g}}  \right) - \bm{e}_{U^2-1}\left( \frac{\phi(\chi \beta_{gu} - \alpha_{g}) \sum_{l=1}^{U} \frac{1}{\zeta_{gl}}}{ \chi \rho \beta_{gu} \alpha_{g}}\right)  \right.$} \nonumber\\
    &\text{\small $\times e^{- \frac{\phi(\chi \beta_{gu} - \alpha_{g}) \sum_{l=1}^{U} \frac{1}{\zeta_{gl}}}{ \chi \rho \beta_{gu} \alpha_{g}} } \mathrm{Ei}\left(\hspace{-1mm} - \frac{\phi \sum_{l=1}^{U} \hspace{-1mm} \frac{1}{\zeta_{gl}}}{\rho \chi \beta_{gu} } \hspace{-1mm} \right) \hspace{-1mm} + \hspace{-.6mm} e^{-\frac{\phi \sum_{l=1}^{U} \frac{1}{\zeta_{gl}} }{\rho \alpha_{g}} } \hspace{-1mm} \sum_{m = 1}^{U^2 - 1} \hspace{-1mm} \frac{1}{m!}$} \nonumber\\ 
    & \text{\small $\left. \times \left( - \frac{\chi \beta_{gu} - \alpha_{g} }{\alpha_{g} }\right)^{m} \sum_{k = 0}^{m-1} (m-k-1)! \left(\hspace{-1mm} - \frac{\phi \sum_{l=1}^{U} \hspace{-1mm} \frac{1}{\zeta_{gl}}}{\rho \chi \beta_{gu}} \hspace{-1mm} \right)^K \right]\hspace{-1mm} .$} \hspace{-1mm}
\end{align}

\textit{Proof:} Please, see Appendix \ref{ap6}.  \hfill $\blacksquare$

\paragraph*{Proposition IV} When the BS employs the proposed dual-polarized MIMO-RSMA scheme, the ergodic sum-rate for the private message experienced in the $g$th group can be approximated by
\begin{align}\label{erg_pri_a1}
    C^p_g &=  \sum_{u=1}^{U} \frac{ \beta_{gu}}{ \mathrm{ln}(2)(\beta_{gu} - \chi \alpha_{g})} \left[ e^{\frac{\phi}{\rho \zeta_{gu}\chi\alpha_{g}}} \mathrm{Ei}\left(- \frac{ \phi}{\rho \zeta_{gu}\chi\alpha_{g}} \right) \right. \nonumber\\
    &\left. - e^{\frac{\phi}{\rho \zeta_{gu}\beta_{gu} }} \mathrm{Ei}\left(- \frac{ \phi}{\rho \zeta_{gu} \beta_{gu} } \right)  \right].
\end{align}

\textit{Proof:} Please, see Appendix \ref{ap7}.  \hfill $\blacksquare$

\section{Numerical and Simulation Results}\label{si_res_sec}
This section validates the theoretical analysis and demonstrates the performance superiority of the dual-polarized MIMO-RSMA strategy over conventional single-polarized and dual-polarized baseline systems, including MIMO-OMA, which implements time-division multiple access, single-polarized MIMO-RSMA, single-polarized MIMO-NOMA, and the dual-polarized MIMO-NOMA scheme from \cite{ni3}.


The BS of the dual-polarized schemes employs $\frac{M}{2} = 50$ co-located pairs of dual-polarized antennas, i.e., a total of $M=100$ transmit antennas. For fair performance comparisons, the single-polarized systems implement the same number of antennas, and the covariance matrices for all systems are generated using the one-ring model \cite{ref1}. We assume that users are distributed among $G = 4$ spatial groups, with the $g$th group positioned at the azimuth angle given by $\theta_{g} = 30^\circ + (g-1) 160^\circ$ and a distance of $170$~m from the BS to its center.
The presented results are based on the first group, which is located at the azimuth angle of $30^\circ$. Within each group there are $U=3$ users, in which users $1$, $2$, and $3$ are located, respectively, at $d_1 = 200$~m, $d_2 = 170$~m, and $d_3 = 140$~m from the BS. Under this setup, the large-scale fading coefficient for the $u$th user is modeled by $\zeta_{u} = \delta d_{u}^{-\eta}$, where $\delta$ is a BS array gain parameter that is adjusted based on the desired users' performance, and $\eta$ is the path-loss exponent. These parameters are configured as $\delta  = 4 \times 10^4$ and $\eta = 2.5$. Moreover, we set $\bar{M} = 6$ and adopt a fixed power allocation, where, for the RSMA schemes, we adjust $\alpha = 0.7$ and $\beta_u = (1 - \alpha)/U = 0.1$, for $u=1,\cdots, 3$. For the NOMA schemes, the power coefficients of users $1, 2$, and $3$ are set to $5/8, 2/8$, and $1/8$, respectively, whereas for OMA, the BS transmits at each time-slot using its full power.

\begin{figure}[t]
	\centering
	\includegraphics[width=1\linewidth]{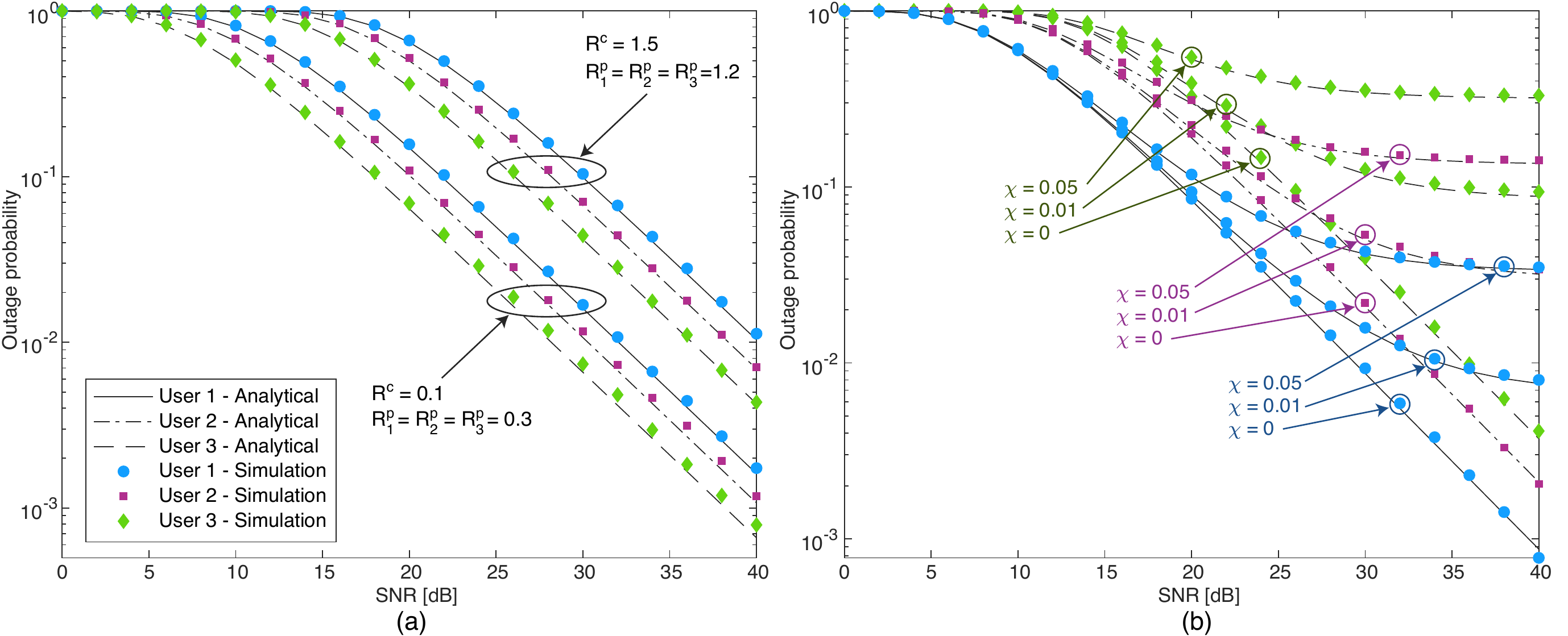}
	\caption{Analytical and simulated outage probabilities: (a) for different values of target rates with $\chi = 0$, and (b) for different values of $\chi$ with $R^c = 0.5$~bpcu, $R^p_{1} = 0.1$~bpcu, $R^p_{2} = 0.5$~bpcu, and $R^p_{3} = 1.2$~bpcu.}\label{f2}
\end{figure}

Figs. \ref{f2}(a) and \ref{f2}(b) compare the analytical and simulated outage probabilities achieved with the proposed dual-polarized MIMO-RSMA scheme. Fig. \ref{f2}(a), specifically, shows outage probability curves for different sets of target rates considering a scenario with negligible cross-polar interference, i.e., for $\chi = 0$. It can be seen in this scenario that, independently of the specified target rates, the approximate analytical curves generated with \eqref{out_comm_a1} and \eqref{out_private_a1} can follow the simulated ones with high accuracy. In Fig. \ref{f2}(b), the performance of the dual-polarized MIMO-RSMA is tested for different levels of cross-polar interference, and an accurate agreement between simulated and analytical results can also be observed.

\begin{figure}[t]
	\centering
	\includegraphics[width=1\linewidth]{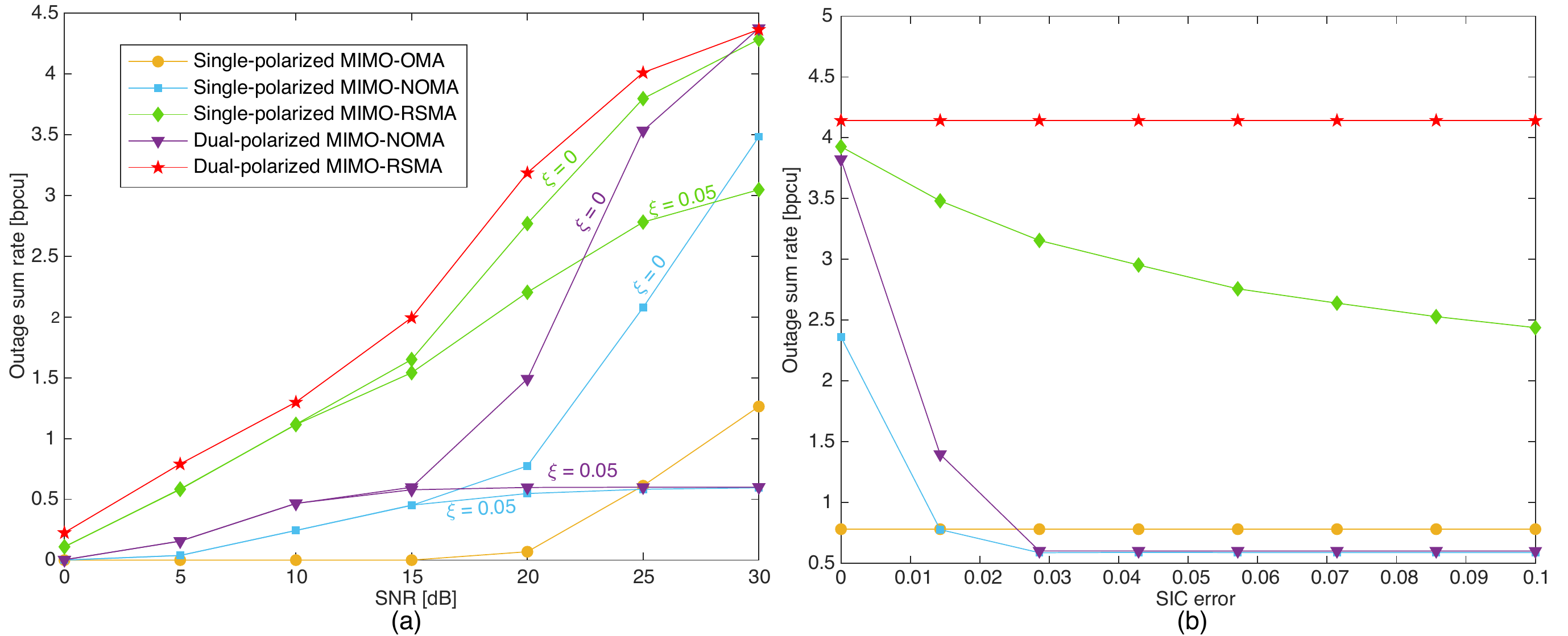}
	\caption{Outage sum-rate curves: (a) versus SNR with different values of $\xi$, and (b) versus $\xi$ for a fixed SNR value of $24$~dB. ($R^c = 0.5, R^p_{1} = 0.1, R^p_{2} = 1, R^p_{3} = 2$~bpcu, $\chi = 0.001$).\vspace{-4mm}}\label{f4}
\end{figure}

Figs. \ref{f4}(a) and \ref{f4}(b) present the outage sum-rates achieved by the proposed dual-polarized MIMO-RSMA strategy and by other conventional MA systems. For a fair comparison, the target rates for schemes other than RSMA are adjusted as $R^c + R^p_u$. 
Fig. \ref{f4}(a) reveals the performance superiority of the proposed dual-polarized MIMO-RSMA scheme over the baseline systems in scenarios with and without residual SIC errors. The effects of increasing the SIC error factor on the outage sum-rates can be observed in Fig. \ref{f4}(b). As can be seen, the MIMO-NOMA systems rapidly become less spectrally efficient than MIMO-OMA with the increase of $\xi$. On the other hand, the dual-polarized MIMO-RSMA scheme show robustness to imperfect SIC.

%

\begin{figure}[t]
	\centering
	\includegraphics[width=1\linewidth]{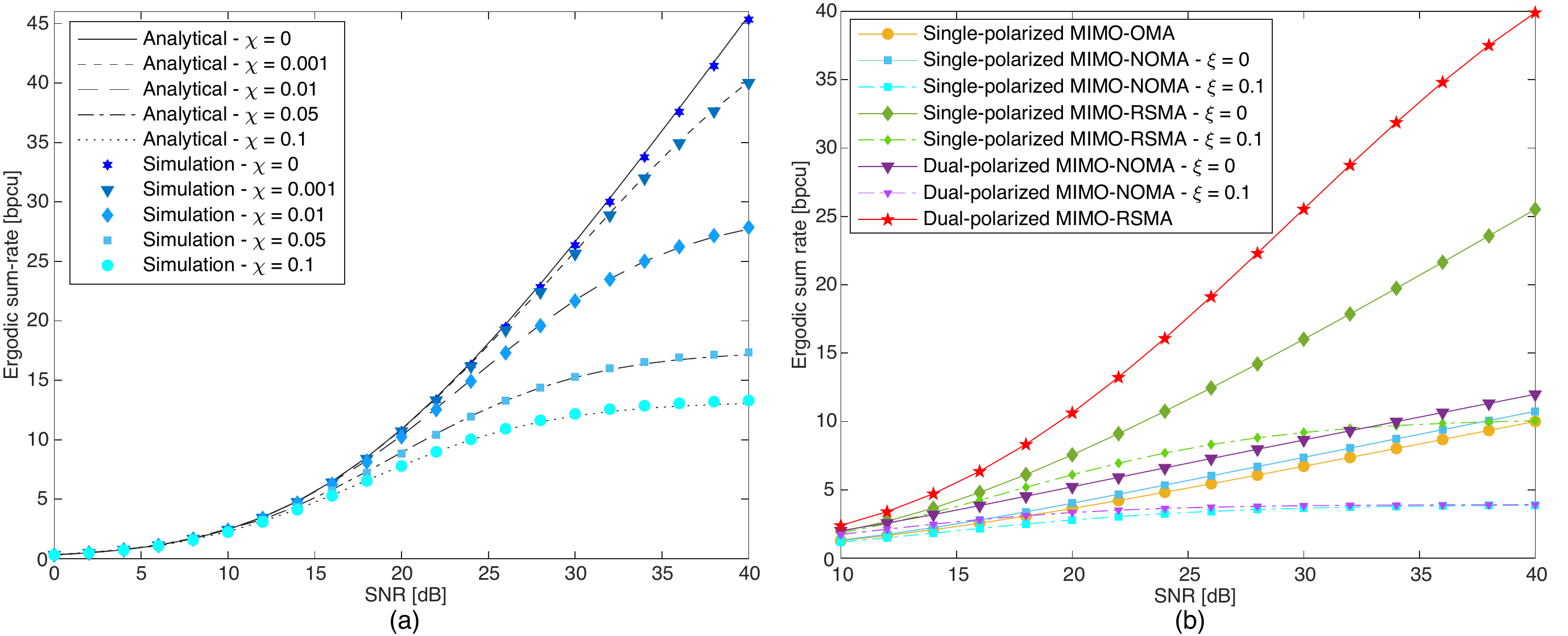}
	\caption{(a) analytical and simulated ergodic sum-rates of dual-polarized MIMO-RSMA for various values of $\chi$, and (b) simulated ergodic sum-rates of various MA schemes for different values of $\xi$ with $\chi = 0.001$.}\label{f5}
\end{figure}

The analytical ergodic sum-rate is validated in Fig. \ref{f5}(a). As can be observed, a near-perfect agreement between analytical and simulated curves can be obtained for all values of $\chi$.
Last, Fig. \ref{f5}(b) shows simulated ergodic sum-rates achieved with different schemes. As we can see, the proposed dual-polarized MIMO-RSMA can outperform all baseline schemes. For instance, for an SNR value of $26$~dB, in the scenario with perfect SIC, the dual-polarized MIMO-RSMA achieves a remarkable sum-rate of $19.14$~bpcu, against only $7.29$~bpcu observed for the dual-polarized MIMO-NOMA. 


\vspace{-1mm}

\section{Conclusions}
We have proposed a novel MA strategy for overcoming interference issues of SIC via the polarization domain.
Simulation results validated the theoretical analysis and revealed the performance improvements enabled by that the proposed dual-polarized MIMO-RSMA. It has also been demonstrated that the MIMO-RSMA schemes can impressively outperform conventional single and dual-polarized baseline systems.
\vspace{-1mm}

\section*{Acknowledgments}
This work was partly supported by the Academy of Finland via: (a) FIREMAN consortium n.326270 as part of CHIST-ERA grant CHIST-ERA-17-BDSI-003,  (b) EnergyNet Fellowship n.321265/n.328869/n.352654, and (c) X-SDEN project n.349965.


\appendices

\section{Proof of Proposition I}\label{ap1}
\renewcommand{\theequation}{A-\arabic{equation}}
\setcounter{equation}{0}

The outage probability for the common message can be calculated by
\begin{align}\label{prob_interm}
    P^c_{gu} 
    &= \int_{0}^{\infty} \int_{0}^{\tau_g^c(y + \sigma^2)} f_{\varsigma^c_{gu} \omega^c_{gu}}(x,y) dx dy,
\end{align}
where $\tau_g^c = 2^{R^{c}_{g}} - 1$, and $f_{\varsigma^c_{gu} \omega^c_{gu}}(x,y)$ denotes the joint PDF of $\varsigma^c_{gu}$ and $\omega^c_{gu}$. Given that $\varsigma^c_{gu}$ and $\omega^c_{gu}$ are correlated, deriving their exact joint PDF leads to an intractable mathematical analysis. As an alternative, we assume that $\varsigma^c_{gu}$ and $\omega^c_{gu}$ are independent RVs.
From Section \ref{statc_a1}, we know that $\varsigma^c_{gu}$ follows a Gamma distribution with shape parameter $1$ and rate parameter $\phi/\zeta_{gu}\alpha_{g}$, and $\omega^c_{gu}$ follows a Gamma distribution with shape parameter $U$ and rate parameter $\phi/\zeta_{gu}\chi \beta_{gu}$. Therefore, by recalling  \cite[eq. (15.1.1)]{Krishnamoorthy2006}, the following is achieved
\begin{align}\label{deriv1}
    P^c_{gu} &= \text{\small $\frac{\phi^{U} }{\chi^U\Gamma(U) \zeta_{gu}^{U}\beta_{gu}^U}  \left( \int_{0}^{\infty} y^{U-1} e^{-y\frac{\phi}{ \zeta_{gu} \chi \beta_{gu}}} dy  \right.$}\nonumber\\
    & \text{\small $- \left. e^{-\frac{\phi \tau_g^c \sigma^2}{\zeta_{gu} \alpha_{g}}} \int_{0}^{\infty} y^{U-1} e^{-y\left(\frac{\phi}{ \zeta_{gu} \chi \beta_{gu}} + \frac{\phi \tau_g^c}{\zeta_{gu} \alpha_{g}} \right)}  dy \right)$}.
\end{align}
The two integrals in \eqref{deriv1} are of the form $\int_{0}^{\infty} y^{a-1} e^{-yb} dy$, which has a solution given by $b^{-a}\Gamma(a)$ \cite[eq. (3.381.4)]{ref8}. Therefore, by defining $\rho = \sigma_n^{-2}$, \eqref{deriv1} can be solved as
\begin{align}\label{pout_ap1}
    P^c_{gu} &= 1 - \left(\frac{\alpha_{g}}{\alpha_{g} + \chi\beta_{gu}\tau_g^c } \right)^U  e^{-\frac{\phi \tau_g^c}{\rho \zeta_{gu} \alpha_{g}}},
\end{align}
which completes the proof. \hfill $\blacksquare$

\section{Proof of Proposition II}\label{ap2}
\renewcommand{\theequation}{B-\arabic{equation}}
\setcounter{equation}{0}

Recall from Section \ref{statc_a1} that $\varsigma^p_{gu}$ and $\omega^p_{gu}$ follow Gamma distributions with shape parameters 1 and rate parameters $\phi / \zeta_{gu} \beta_{gu}$ and $\phi/\zeta_{gu}\chi\alpha_{g}$, respectively. Given this, and assuming that $\varsigma^p_{gu}$ and $\omega^p_{gu}$ are independent, their joint PDF can be obtained by
\begin{align}\label{jointpdf2}
    f_{\varsigma^p_{gu} \omega^p_{gu}}(x,y) 
    &= \frac{\phi^2}{\zeta_{gu}^2 \chi\alpha_{g} \beta_{gu}} e^{-y\frac{\phi}{\zeta_{gu}\chi\alpha_{g}}} e^{-x\frac{\phi}{\zeta_{gu} \beta_{gu}}}.
\end{align}

Then, with the SINR in \eqref{sinr_a1_p}, and defining $\tau_{gu}^p = 2^{R^{p}_{gu}} - 1$, the outage probability for the private massages can be derived as follows
\begin{align}\label{probp2}
    P^p_{gu} & = \frac{\phi}{\zeta_{gu}\chi\alpha_{g}}  \left( \int_{0}^{\infty} e^{-y\frac{\phi}{\zeta_{gu}\chi\alpha_{g}}} dy \right. \nonumber\\
    &- \left. e^{-\frac{\phi}{\zeta_{gu} \beta_{gu}} \tau_{gu}^p\sigma^2} \int_{0}^{\infty} e^{-y\left( \frac{\phi}{\zeta_{gu} \beta_{gu}} \tau_{gu}^p + \frac{\phi}{\zeta_{gu}\chi\alpha_{g}} \right)}  dy \right) \nonumber\\
    & = 1 - \frac{\beta_{gu}}{ \chi\alpha_{g}\tau_{gu}^p + \beta_{gu}} e^{-\frac{\phi \tau_{gu}^p}{\rho\zeta_{gu}\beta_{gu}}},
\end{align}
which completes the proof. \hfill $\blacksquare$

\section{Proof of Proposition III}\label{ap6}
\renewcommand{\theequation}{C-\arabic{equation}}
\setcounter{equation}{0}
First, we need to obtain the CDF of $\min \{\gamma^c_{gu}\}$, for $u=1, \cdots, U$, denoted by $F_{\hspace{-1mm}\text{\tiny min$\gamma^c_{gu}$}}(z)$. From Appendix \ref{ap1}, we can easily achieve the CDF of $\gamma^c_{gu}$ by replacing $\tau^c_{gu}$ by $z$ in the outage probability expression in \eqref{pout_ap1}. In view of this, we can exploit the theory of Order Statistics to calculate $F_{\hspace{-1mm}\text{\tiny min$\gamma^c_{gu}$}}(z)$. Specifically, by knowing that $\gamma^c_{g1}, \cdots, \gamma^c_{gU}$ are not identically distributed, we can recall \cite[eq. (5.4.11)]{ref7} to obtain the desired CDF as follows
\begin{align}
    F_{\hspace{-1mm}\text{\tiny min$\gamma^c_{gu}$}}(z) &= 1 - \prod_{l = 1}^{U} \left(\frac{\alpha_{g}}{\alpha_{g} + \chi\beta_{gu}z } \right)^U  e^{-z\frac{\phi}{\rho\zeta_{gl}\alpha_{g}}} \nonumber\\
    &= 1 - \alpha_{g}^{U^2} \left({\alpha_{g} + \chi\beta_{gu}z } \right)^{-U^2}  e^{-z\frac{\phi \sum_{l=1}^{U} \frac{1}{\zeta_{gl}}}{\rho\alpha_{g}}}.
\end{align}

Now that we know $F_{\hspace{-1mm}\text{\tiny min$\gamma^c_{gu}$}}(z)$, the desired sum-rate can be obtained by a Riemann-Stieltjes integral, as follows
\begin{align}
    C^c_{g} &= \sum_{u=1}^{U} \mathrm{E}\left\{ \min_{l\in \{1, \cdots, U\} }\left\{ \log_2\left(1 + \gamma^c_{gl}\right) \right\} \right\} \nonumber\\
    &= \sum_{u=1}^{U} \int_{-\infty}^{\infty} \log_2\left(1 + z \right) dF_{\hspace{-1mm}\text{\tiny min$\gamma^c_{gu}$}}(z). 
\end{align}
Next, by integrating by parts and applying the transformation $\alpha_{g} + \chi\beta_{gu} z = t$, we achieve
\begin{align}\label{preerg_ap6}
    C^c_{g} 
    & = \sum_{u=1}^{U} \frac{\alpha^{U^2}_{g}}{\mathrm{ln}(2)} \left[ - \alpha_{g}^{-U^2} e^{\frac{\phi \sum_{l=1}^{U} \frac{1}{\zeta_{gl}} }{\rho \alpha_{g}} } \mathrm{Ei}\left(- \frac{\phi \sum_{l=1}^{U} \frac{1}{\zeta_{gl}} }{\rho \alpha_{g}} \right) \right. \nonumber\\
    & + U^2 e^{\frac{\phi \sum_{l=1}^{U} \frac{1}{\zeta_{gl}} }{\rho \alpha_{g}} } \int_{\alpha_{g}}^{\infty}  t^{-U^2 - 1} \mathrm{Ei}\left(-\left(\frac{\chi\beta_{gu} - \alpha_{g}}{\chi\beta_{gu}} \right. \right. \nonumber\\
    & \left. + \left. \left. \frac{t}{\chi\beta_{gu}} \right)\frac{\phi \sum_{l=1}^{U} \frac{1}{\zeta_{gl}}}{\rho\alpha_{g}} \right) dt \right].
\end{align}\vspace{1mm}

For solving the integral in \eqref{preerg_ap6}, we make use of \cite[eq. (4.1.24)]{Geller69}, in which, after some simplifications, we can finally obtain the desired ergodic sum-rate expression in \eqref{erg_comm_a1}, which completes the proof. \hfill $\blacksquare$

\section{Proof of Proposition IV}\label{ap7}
\renewcommand{\theequation}{D-\arabic{equation}}
\setcounter{equation}{0}
For deriving the ergodic sum-rate of the private messages, we obtain the CDF of $\gamma^p_{gu}$ from the outage probability in \eqref{probp2}. Then, similarly as in Appendix \ref{ap6}, the desired sum-rate is calculated by a Riemann-Stieltjes integral, as follows 
\begin{align}
    C^p_{g} &= \hspace{-1mm} \sum_{u=1}^{U} \frac{\beta_{gu}}{\mathrm{ln}(2)}\int_{0}^{\infty} \hspace{-2mm} (1 + z)^{-1}(\chi\alpha_{g}z + \beta_{gu})^{-1} e^{-z\frac{\phi}{\rho\zeta_{gu}\beta_{gu}}} dz.
\end{align}
By integrating by parts and applying the transformation $\chi\alpha_{g}z + \beta_{gu} = t$, we get
\begin{align}
    C^p_{g} 
    %
    &= \sum_{u=1}^{U} \frac{\beta_{gu}}{\mathrm{ln}(2)} \left[ \beta_{gu}^{-1} e^{\frac{\phi}{\rho\zeta_{gu}\beta_{gu}}} \mathrm{Ei}\left(-\frac{\phi}{\rho\zeta_{gu}\beta_{gu}} \right)  \right. \nonumber\\
    & + e^{\frac{\phi}{\rho\zeta_{gu}\beta_{gu}}} \int_{\beta_{gu}}^{\infty}  t^{-2} \mathrm{Ei}\left(-\left(\frac{t}{\chi \alpha_{g}} \right. \right. \nonumber\\
    &+ \left. \left. \left. \frac{\chi \alpha_{g} - \beta_{gu}}{\chi \alpha_{g} } \right)\frac{\phi}{\rho\zeta_{gu}\beta_{gu}} \right) dt \right].
\end{align}\vspace{1mm}

Finally, by recalling \cite[eq. (4.1.21)]{Geller69}, and performing some algebraic manipulations, the ergodic sum-rate of the private messages for the $g$th group can be obtained as in \eqref{erg_pri_a1}, which completes the proof. \hfill $\blacksquare$

\ifCLASSOPTIONcaptionsoff
\newpage
\fi

\bibliographystyle{IEEEtran}
\bibliography{output}

\end{document}